\documentclass[twocolumn,showpacs,preprintnumbers,amsmath,amssymb]{revtex4}

\usepackage{graphicx}
\usepackage{dcolumn}
\usepackage{bm}

\begin{document}


\title{Construction of localized wave functions for a disordered optical lattice and analysis of the resulting
Hubbard model parameters}

\author{S. Q. Zhou}
\author{D. M. Ceperley}%
 \email{ceperley@illinois.edu}
\affiliation{%
Department of Physics, University of Illinois at Urbana-Champaign, Urbana, Illinois 61801, USA
}

\date{\today}

\begin{abstract}
We propose a method to construct localized single particle wave
functions using imaginary time projection and thereby determine lattice Hamiltonian
parameters. We apply the method
to a specific disordered potential generated by an optical
lattice experiment and calculate for each instance of disorder,
the equivalent lattice model parameters. The probability
distributions of the Hubbard parameters are then determined.
Tests of localization and eigen-energy convergence are examined.
\end{abstract}

\pacs{Valid PACS appear here}
\maketitle

\section{\label{sec:level1}INTRODUCTION}
Understanding the properties of disordered materials has a
fundamental significance in condensed matter physics. Various
kinds of disorder exist in real materials, but their disorder
is difficult to characterize and control experimentally.
Recently developed optical lattice techniques \cite{demarco}
have enabled the construction of a nearly perfectly controlled
disordered potential and the measurement of properties of
strongly correlated atoms in that potential provides an
opportunity to compare quantitatively experimental results with
parameter-free theoretical calculations. Bosonic atoms are
particularly interesting because then Monte Carlo simulation is
efficient.

In this paper, we consider the problem of mapping a disordered
single body potential to a lattice model. By removing the high
energy states associated with the continuum, the Monte Carlo
simulation becomes more efficient. Particularly, efficient algorithms
have been developed \cite{sse}\cite{worm} for lattice models.
Consider the continuum Hamiltonian of $N$ atoms with mass $m$
moving in the external potential $U(\mathbf{r})$ and
interacting with the pairwise potential energy $V({\mathbf{r}}_{\alpha}-{\mathbf{r}}_{\beta})$
\begin{eqnarray}
  \hat{{\mathcal{H}}}_{N} &=& \sum_{\alpha=1}^{N} \left[ \frac{{\mathbf{p}}_{\alpha}^{2} }{2m}  + U({\mathbf{r}}_{\alpha})\right] + \sum_{\alpha<\beta}V({\mathbf{r}}_{\alpha} - {\mathbf{r}}_{\beta})
\label{eq:one}
\end{eqnarray}
where indices $\alpha$, $\beta$ label the atoms. On the other hand, the
quantum mechanics of particles moving in a lattice is
conveniently described in a basis of localized wave functions,
such as the Wannier functions associated with a periodic
potential. Using these localized functions, we can define an
effective lattice Hubbard Hamiltonian.  Written in second
quantized notation it has the form:
\begin{eqnarray}
\nonumber \hat{h}  &=&  -\sum_{\langle ij\rangle}t_{ij}
a_{i}^{\dag}a_{j} + \sum_{i}\epsilon_{i}n_{i} +
\frac{1}{2}\sum_{i}u_{i}n_{i}(n_{i}-1) \\ && -
\sum_{\{ij\}}\tilde{t}_{ij} a_{i}^{\dag}a_{j} +
\frac{1}{2}\sum_{\langle ij\rangle}\tilde{u}_{ij}n_{i}n_{j} +
\cdots\cdots \label{eq:two}
\end{eqnarray}
where $i$ labels the single particle states (lattice
sites), $\langle ij\rangle$ denotes a nearest neighbor pair and
$\{ij\}$ a next-nearest neighbor pair, $t_{ij}$ and $\tilde{t}_{ij}$ are
hopping coefficients, $\epsilon_{i}$ is the on-site energy, $u_{i}$ is
the on-site interaction and $\tilde{u}_{ij}$ is the nearest neighbor
off-site interaction. (Terms such as next-nearest neighbor
hopping and offsite interaction are often neglected.) Note that
${\mathcal{\hat{H}}}_{N}$ refers to the $N$-body Hamiltonian in
continuous space and $\hat{h}$ to its equivalent on a lattice.

In a periodic potential, Wannier functions of a given band are
related to the Bloch functions $\psi_{n{\mathbf{k}}}$ of the same
band $n$ by the unitary transformation
\begin{eqnarray}
  w_{ni}({\mathbf{r}}) = w_{n}({\mathbf{r}}-{\mathbf{R}}_{i}) &=& \frac{1}{\sqrt{N}} \sum_{{\mathbf{k}}} \psi_{n{\mathbf{k}}}({\mathbf{r}}) e^{-i{\mathbf{k}}\cdot {\mathbf{R}}_{i}}.
\end{eqnarray}
$w_{ni}$ is localized around the lattice site
${\mathbf{R}}_{i}$ \cite{wannier}. However, in the absence of
periodicity, the concept of Wannier functions needs to be
generalized. Two main types of generalizations exist in
literature. The perturbative approach
\cite{pert1} assumes the existence of
the band structure and thus is applicable to nearly periodic
potentials. The variational approach
\cite{max1}\cite{max2}\cite{max3} emphasizes the minimization
of the spatial spread with respect to unitary transformations
of a starting basis set, for example the Wannier functions of a
periodic potential.

In order to be useful, we would like the generalized Wannier
functions to have the following properties. First, localization
is required by the physical picture of particles hopping in the
lattice. Second, a correct description of the low energy density
of states is necessary to capture the low temperature physics.
Third, for convenience, the orthogonality of the basis set is
required to use commutation relations of creation and
annihilation operators in the second quantized Hamiltonian.
Finally, we would like the lattice Hamiltonian to be free of
the sign problem so that quantum Monte Carlo calculations are
efficient. This requires the off-diagonal elements to be
non-positive(\emph{i.e.} $t_{ij}\ge 0$). Note that the original
Hamiltonian $\hat{{\mathcal{H}}}_{N}$ has this property.

In section II, we propose a method of constructing localized
single particle basis functions based on imaginary time
evolution of localized basis functions: $w_i(0)$ where $i$
labels the site.
\begin{eqnarray}
\left| w_i (\tau) \right\rangle  &\equiv& e^{-\tau
\hat{{\mathcal{H}}}_{1}} \left| w_i (0) \right\rangle
\end{eqnarray}
where $\hat{{\mathcal{H}}}_{1}$ denotes the one particle continuum
Hamiltonian. This has the effect of suppressing the high energy
components but also spreading out the basis states. In section III,
results of the method are presented for the specific disorder probed
experimentally.

\section{Method}
In constructing a lattice model, our goal is to coarse grain the
description of the continuum
system, so that instead of recording the precise position of an
atom, we only record which lattice site it occupies. We match up
 the lattice and continuum models using the density matrix; we
require that the low temperature density matrix of the lattice model to be
identical to the reduced density matrix of the continuum system when
high energy degrees of freedoms are traced out.
Use of the density matrix is motivated by the fact that the linear response of a
system to an external perturbation, either an external field,
or particle insertion, or a coupling to another subsystem is
determined by its one-body density matrix \cite{ceperley}\cite{Feynman}.
If we match the density matrices, the lattice system is guaranteed to have
not only the same density distribution $n({\mathbf{r}})$ and hopping properties,
such as diffusion, but also the same response to external
perturbations as the continuum system.

The unnormalized single particle density matrix in the continuum system is
defined by:
\begin{equation}
\rho\left({\mathbf{r}},{\mathbf{r'}} ;\tau\right)=	\left
\langle {\mathbf{r}} | e^{-\tau{\hat{\mathcal{H}}}_{1} } |
{\mathbf{r'}} \right\rangle.
\label{eq:rho}
\end{equation}
Let $w_i(\mathbf{r};0)$ be a localized basis which assigns
atoms to lattice sites, e.g. $w_i ({\mathbf{r}};0) =1$ if
$|\mathbf{r}-{\textbf{R}}_i|$ is minimized with respect to $i$, i.e. in
the $i^{th}$ Wigner-Seitz cell.  Then a course-grained density
matrix is defined as
\begin{eqnarray}
\nonumber  S_{ij} (\tau) &=& \left\langle w_{i}(0) |e^{-\tau\hat{{\mathcal{H}}}_{1} }|w_{j}(0) \right\rangle \\
   &=& \int d{\mathbf{r}} d{\mathbf{r}}'  w_i^*({\mathbf{r}};0) \rho( {\mathbf{r}},{\mathbf{r}}'; \tau) w_j
({\mathbf{r}}';0)
\label{eq:sij}
\end{eqnarray}
Note that if $w_i(\mathbf{r};0)$'s are chosen to be everywhere positive, all elements of the
lattice density matrix $S_{ij}$ are also positive and can be
used in a lattice QMC calculation directly. We now want to construct a single-particle Hamiltonian which
when solved gives $S _{ij} (\tau)$ for large $\tau$, or in
matrix notation to determine $\hat{h}$ such that
\begin{equation}
    \hat{S} (\tau) = e^{-\tau \hat{h}}.
\label{eq:submat}
\end{equation}
\footnote{$S$ is Hermitian and positive definite, so $h$ exists uniquely.} Formally,
the solution $\hat{h} = -\tau^{-1}\log \hat{S}\left(\tau\right)$ may have some $\tau$ dependance and not
necessarily have the other properties mentioned above.
Differentiating Eq.~(\ref{eq:submat}) with
respect to $\tau$ and multiply on the right and left by $\hat{S}^{-1/2}$, we find an expression for $h$ in
terms of $S$:
\begin{equation}
    \hat{h} = -\hat{S}^{-\frac{1}{2}} \frac{d\hat{S}}{d\tau}\hat{S}^{-\frac{1}{2}}
 -\int_{0}^{\tau} e^{\left( \frac{\tau}{2}-\lambda \right)\hat{h}} \left( \frac{d\hat{h}}{d\tau}\right)
 e^{\left( \lambda - \frac{\tau}{2} \right)\hat{h}} d\lambda.
\label{eq:latham}
\end{equation}
If we assume that $h$ becomes $\tau$-independent as $\tau\rightarrow\infty$, we can neglect the second
term on the right hand side and find:
\begin{equation}
    \hat{h} = -\hat{S}^{-\frac{1}{2}} \frac{d\hat{S}}{d\tau}\hat{S}^{-\frac{1}{2}}.
\label{eq:latham2}
\end{equation}
Consider the eigenfunction expansion of the continuum
density matrix:
 \begin{equation}
 \rho({\mathbf{r}},{\mathbf{r}}';\tau)= \sum_{\alpha}
 \phi_{\alpha}^*({\mathbf{r}})\phi_{\alpha}({\mathbf{r}}')e^{-\tau E_{\alpha}}
 \end{equation}
where $E_{\alpha}$ and $\phi_{\alpha}$ are the $1$-particle eigenvalues
and eigenfunctions of the continuum Hamiltonian. For a
sufficiently large $\tau$, and for a system with a gap, only
states below the gap will survive. If there are $N$ such
states, it is clear that we will capture the density of states
with exactly $N$ basis  functions $w_{i}$.
Now let us define the orthogonalized basis by splitting up the density operator
\begin{equation}
    \exp\left(-\tau \hat{{\mathcal{H}}}_{1}\right) = \exp\left( -\frac{1}{2}\tau \hat{{\mathcal{H}}}_{1}\right)\exp\left( -\frac{1}{2}\tau \hat{{\mathcal{H}}}_{1}\right)
\end{equation}
and having it act partially to the left and right in Eq.~(\ref{eq:sij}).
Combining Eq.~(\ref{eq:sij}) and Eq.(\ref{eq:latham2}) we obtain the expression for the
model Hamiltonian:
\begin{eqnarray}
\nonumber  h_{ij} &=& \sum_{kl} S^{-\frac{1}{2}}_{ik}(\tau) \left\langle w_{k} (\tau/2) |\hat{{\mathcal{H}}}_{1}| w_{l} (\tau/2) \right\rangle S^{-\frac{1}{2}}_{lj}(\tau) \\
   &=&  \left\langle \tilde{w}_{i}(\tau/2)| \hat{{\mathcal{H}}}_{1}| \tilde{w}_{j}(\tau/2) \right\rangle
\label{eq:latham3}
\end{eqnarray}
where $w_{i}(\tau)=e^{-\tau \hat{{\mathcal{H}}}_{1}} w_{i}(0)$ are the
non-orthonormalized basis functions at time $\tau$ and
$\tilde{w}(\tau/2)=\hat{S}^{-1/2}(\tau) w (\tau/2)$ are the
orthonormalized basis functions because
\begin{equation}
    S_{ij} (\tau) = \left\langle w_{i} (\tau/2) | w_{j} (\tau/2) \right\rangle
\end{equation}
is the overlap matrix. This is known as
\emph{L\"{o}wdin orthogonalization}
\cite{lowdin1}\footnote{Mathematically equivalent to the
procedure for periodic potentials introduced by
Wannier\cite{wannier}}.

The imaginary time evolution is equivalent to a diffusion
process with sinks or sources determined by the potential
$U({\mathbf{r}})$ . Without a potential present, an initially localized
distribution will spread out as $\sqrt{\tau}$ as a function of
imaginary time. When the wavepacket (or basis function) $\left| w_i (\tau) \right\rangle  \equiv e^{-\tau
\hat{{\mathcal{H}}}_{1}} \left| w_i (0) \right\rangle$
encounters the regions of high potential energy separating
the lattice sites, its diffusion will stop, until it tunnels
through to the next site. If the assumption of temperature-independence
\begin{equation}
    \lim_{\tau\rightarrow\infty} \left(\frac{d\hat{h}}{d\tau}\right) = 0
\end{equation}
is correct, according to Eq.~(\ref{eq:latham3}), the orthogonalized basis
$\hat{S}^{-1/2}\left(2\tau\right)\left| w_i (\tau) \right\rangle$ converges
at large $\tau$. Instead of taking the logarithm of the reduced density matrix
Eq.~(\ref{eq:submat}), we choose to construct the lattice Hamiltonian from
Eq.~(\ref{eq:latham2}) for two reasons. Firstly, numerical tests show that
Eq.~(\ref{eq:latham2}) converges faster than $-\frac{1}{\tau}\log \hat{S}$
as $\tau$ increases. Secondly, the explicit construction of basis functions enables
us to calculate the interaction terms in the second quantized many body Hamiltonian.
Finally, use of Eq.~(\ref{eq:latham2}) instead of Eq.~(\ref{eq:submat}) gives a spectrum
as upper bounds to the continuum spectrum.

The choice of the initial basis function $w_{i}({\mathbf{r}};0)$ is to
some  extent arbitrary, as long as it is non-negative and localized within a
lattice cell, and there is one basis function for each lattice site. We choose
to set $w_i ({\mathbf{r}};0)=1$ inside a cube of side $\sigma$ centered on ${\mathbf{R}}_{i}$.

\subsection{Algorithm for imaginary time projection and orthogonalization}
To apply the imaginary time evolution to the construction of
localized wave functions and thereby to extract microscopic
parameters of the corresponding lattice model, we then start
with $N$ initial trial wave functions, each of which is well
localized in one lattice cell. Each wave function is
\emph{independently} evolved over a sufficiently long imaginary
time. The set of $N$ wave functions are then transformed into an orthonormal basis.

To perform the imaginary time evolution, consider the Trotter formula\cite{Trotter}
\begin{equation}
e^{-\beta \hat{{\mathcal{H}}}_{1}} = \lim_{n\rightarrow \infty} \left(
e^{-\frac{\beta}{n} \hat{T}}e^{-\frac{\beta}{n} \hat{U}}\right)^{n}.
\end{equation}
In a coordinate representation, a single step of imaginary time
$\tau$ can be written as:
\begin{eqnarray}
\nonumber  w({\mathbf{r}},t+\tau) &=& \int d^{3}{\mathbf{r}}' \langle {\mathbf{r}} |e^{-\tau \hat{{\mathcal{H}}}_{1}}|{\mathbf{r}}' \rangle w({\mathbf{r}}',t) \\
 &=&  \left( \frac{m}{2\pi\hbar\tau} \right)^{3/2}\\
\nonumber && \times \int d^{3}{\mathbf{r}}' e^{-\frac{
m}{2\hbar\tau}\left({\mathbf{r}}' -{\mathbf{r}} \right)^{2}
}e^{-\frac{\tau U({\mathbf{r}}')}{\hbar}}w({\mathbf{r}}',t).
\end{eqnarray}
This integral is a convolution, and  can be efficiently
evaluated by Fast Fourier Transform
\begin{equation}
    w({\mathbf{r}},t+\tau) = {\mathrm{FFT}} \left[ e^{-\frac{\tau \hbar  {\mathbf{k}}^{2}}{2m}}f_{{\mathbf{k}}} \right]
\label{eq:fft}
\end{equation}
where $f_{{\mathbf{k}}}$ is defined as an
inverse-Fourier transform
\begin{equation}
    f_{{\mathbf{k}}} = {\mathrm{FFT}}^{-1}\left[ e^{-\frac{\tau U({\mathbf{r}})}{\hbar}}w({\mathbf{r}},t)  \right].
\end{equation}
We can also take advantage of the localization of
$w({\mathbf{r}})$: it is vanishingly small away from its
initial site, so that we only store its values in a cube
enclosing the region in which the wave function is non-zero.
When doing the second FFT, Eq.~(\ref{eq:fft}), we add  a buffer layer
outside the cube with thickness chosen so that the
localization region of the evolved function over one imaginary
time step does not exceed the cube in which FFT is performed; the
thickness is proportional to $\sqrt{\tau/m}$. We
periodically examine the evolved basis set, to determine if the
cube can be made smaller. A common normalization factor
is required for all basis functions every several steps
to avoid numerical overflow or underflow.

Eq.~(\ref{eq:latham2}) demands orthogonalization of the basis set. L\"{o}wdin orthogonalization preserves, as much as
possible, the localization and symmetry of the original
non-orthogonal basis states. In terms of the overlap matrix,
we construct a set of orthogonalized states
\begin{equation}
    |\tilde{w}_{i} \rangle= \sum_{j}(S^{-1/2})_{ij}|w_{j}\rangle.
\end{equation}
No other set of orthonormal states generated from the space
spanned by the original non-orthogonal set of states resemble
the original set more closely, in the sense of least square,
than do L\"{o}wdin set of states \cite{lowdin2}. Explicitly,
L\"{o}wdin orthogonalization minimizes the expression
\begin{equation}
    \phi(\hat{T}) \equiv \sum_{i=1}^{N} \| \hat{T}w_{i} - w_{i}  \|^{2}
\end{equation}
over all linear transformations $\hat{T}$ which transform the
original non-orthonormal set of states $ |w_{i} \rangle $ into
an orthonormal set of states $ |\tilde{w}_{i} \rangle$
\begin{equation}
    \left\langle \tilde{w}_{i} , \tilde{w}_{j} \right\rangle \equiv
    \left\langle \hat{T}w_{i} , \hat{T}w_{j} \right\rangle = \delta_{ij}.
\end{equation}

For efficiency, this procedure can be done in an iterative
fashion. Because the original non-orthogonal set of wave
functions are localized, the overlap matrix $S_{mn}$ has the
form of the identity matrix plus a small off-diagonal part
\begin{equation}
    S_{ij} = \delta_{ij} + A_{ij}
\end{equation}
where the diagonal elements of $A$ are zero and the
off-diagonal elements $|A_{mn}|\ll 1$. This enables us to
perform L\"{o}wdin orthogonalization iteratively by repeated
application of the approximate inverse square root of the
overlap matrix
\begin{equation}
    (\hat{S}^{-1/2})_{ij} \approx \delta_{ij} - \frac{1}{2}A_{ij}
\end{equation}
to the non-orthogonal basis set by updating the overlap matrix
at each step \cite{projection}. Hence the basis set is
iterated as:
\begin{equation}
\tilde{w}\leftarrow \left({\mathbf{1}}-\frac{1}{2}\hat{A} \right)\tilde{w}
\end{equation}
until convergence is reached, $|\hat{A}| \sim 0$. The
convergence of the overlap matrix to identity matrix is
geometric.

The iterative scheme is efficient for large systems
because the basis sets are sparse. The computation time of this algorithm is linear in
the number of lattice sites, i.e. the complexity is proportional to
\begin{equation}
    \left(\#\, \mathrm{of} \, \mathrm{steps}\right)\cdot M\cdot n\log n
\end{equation}
where $M$ is the number of lattice sites and $n$ is the number of pixels for each basis function.

\subsection{Hubbard parameters}
Once the orthogonalized basis set has been constructed, the
effective lattice Hamiltonian is obtained. For convenience we
drop the $\tau$ dependance. According to Eq.~(\ref{eq:latham3}), the single particle Hubbard
parameters are calculated as detailed in Eq.~(\ref{eq:onsite})-(\ref{eq:hopping}): the on-site
energies
\begin{equation}
    \epsilon_{i} = \int \tilde{w}^{*}_{i}({\mathbf{r}})\mathcal{\hat{H}}_{1}\tilde{w}_{i}({\mathbf{r}})d^{3}{\mathbf{r}},
\label{eq:onsite}
\end{equation}
and the hopping coefficients
\begin{equation}
    t_{ij} = -\int
    \tilde{w}^{*}_{i}({\mathbf{r}})\mathcal{\hat{H}}_{1}\tilde{w}_{j}({\mathbf{r}})d^{3}{\mathbf{r}}.
\label{eq:hopping}
\end{equation}
The interaction term is computed from  first-order perturbation
theory in $V$. In the case of a contact interaction with
 the scattering length $a_{s}$  we find for $u$:
\begin{equation}
    u_{i} = \frac{4\pi a_{s}\hbar^{2}}{m}\int  \left| \tilde{w}_{i}({\mathbf{r}})  \right|^{4}d^{3}{\mathbf{r}},
\end{equation}
and the off-site interaction
\begin{equation}
    \tilde{u}_{ij} = \frac{4\pi a_{s}\hbar^{2}}{m}\int  \left| \tilde{w}_{i}({\mathbf{r}})  \right|^{2}  \left| \tilde{w}_{j}({\mathbf{r}})  \right|^{2}d^{3}{\mathbf{r}}.
\label{eq:offu}
\end{equation}

\begin{figure}
\scalebox{0.3}[0.3]{\includegraphics[0,0][30cm,22cm]{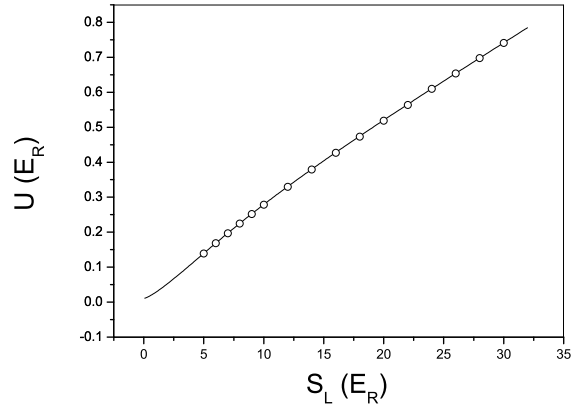}}
\caption{\label{fig:clean1} Hubbard U as a function of the
potential depth of lattice field; the line is obtained from
exact diagonalization, and the open circles
are obtained using the method in this paper.}
\end{figure}

\begin{figure}
\scalebox{0.3}[0.3]{\includegraphics[0,0][30cm,22cm]{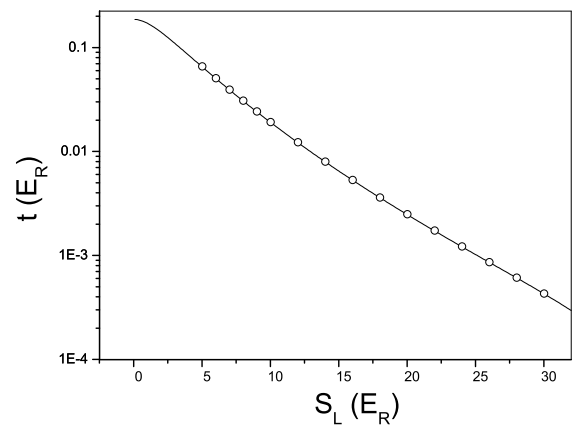}}
\caption{\label{fig:clean2} Log-scale hopping coefficient as a
function of the potential depth of the lattice field; the line
is obtained from exact diagonalization and
the open circles are obtained using the method in this paper.}
\end{figure}

The problem of a single particle moving in a periodic potential
of the form $\cos x + \cos y + \cos z$ can be solved
analytically \cite{exact}. We compared the imaginary time
projected states with the results from exact diagonalization; this is shown in
Fig.~\ref{fig:clean1} and Fig.~\ref{fig:clean2}. We used a spatial grid with $8^{3}$
pixels per lattice cell and an imaginary time step $\Delta \tau = 10^{-4} E_{R}^{-1}$.
We find the error vanishes linearly as the time step goes to zero.

\subsection{Measures of localization and energy}
Several quantities can be used to characterize the localization
and accuracy of the basis set. The spatial spread $\Omega_{w}
\equiv \left\langle r^{2} \right\rangle_{w} - \left\langle
{\mathbf{r}} \right\rangle^{2}_{w}$  quantifies the
localization of a wave function.

The off-site integral $\tilde{u}_{ij}$ measures  the spatial
overlap between a pair of basis states. If it is small relative
to $t_{ij}$ and the Hubbard $U$, the approximation of keeping
only the on-site interaction in the lattice model is
appropriate.  Its rms value over all nearest neighbor pairs
measures the whole basis set.

The convergence rate of the $N\times N$ matrix of the single particle
lattice Hamiltonian is measured by the time derivatives of
its $N$ eigenvalues $E^{(i)}_{\texttt{lattice}}$'s,
\begin{equation}
    \Gamma = \frac{1}{N}\sum_{i}\left| \frac{d}{d\tau} E^{(i)}_{\texttt{lattice}}\right|.
\end{equation}
To determine the accuracy of the basis set we compare $E^{(i)}_{\texttt{lattice}}$'s
with the lowest $N$ eigenvalues $E^{(i)}_{\texttt{exact}}$ of
the original continuum Hamiltonian ${\mathcal{\hat{H}}}_{1}$
estimated from
\begin{equation}
    E^{(i)}_{\texttt{exact}} = E^{(i)}_{\texttt{lattice}}(\tau \rightarrow \infty).
\end{equation}
The \emph{worst case error} is defined as
\begin{equation}
    e_{\texttt{lattice}} \equiv \max_{i} \left| E^{(i)}_{\texttt{lattice}} - E^{(i)}_{\texttt{exact}}\right|.
\label{eq:error1}
\end{equation}
We did the same estimate for the lattice Hamiltonian that has
only nearest neighbor hopping terms: we denote this $e_{\texttt{nn}}$.

\section{Results for a disordered lattice}
We now apply our method to the disordered lattice potential
created in the White et. al. experiment \cite{demarco},
$^{87}$Rb atoms are trapped in a background cubic lattice
potential created by red lasers with wave vector
$k=\frac{\pi}{a}$. The periodic
potential is:
\begin{equation}
    U_{L}({\mathbf{r}}) = -S_{L}\times \sum_{i=1}^{3} \cos\left( \frac{2\pi {\mathbf{n}}_{i}\cdot{\mathbf{r}} }{a} \right)
\end{equation}
where ${\mathbf{n}}_{i}$, $i=1,2,3$ are three mutually
orthogonal unit vectors. A disordered speckle field
$U_{D}({\mathbf{r}})$ is produced by a laser beam with phases
randomized by a diffuser. The speckle field is everywhere positive, characterized by a speckle
strength $S_{D}$:
\begin{eqnarray}
\nonumber  U_{D}({\mathbf{r}}) &>& 0, \quad \forall {\mathbf{r}} \\
\nonumber  \left\langle U_{D}({\mathbf{r}})\right\rangle  &\approx& 0.75 S_{D} \\
\nonumber  \left\langle U_{D}^{2}({\mathbf{r}})\right\rangle  &\approx& S_{D}^{2}
\end{eqnarray}
and a spatial auto-correlation $\left\langle
U_{D}({\mathbf{r}})U_{D}({\mathbf{r}}') \right\rangle$ with correlation length $\sim1.29 a$, i.e. slightly
larger than the lattice spacing.

The total external potential is a superposition of
the periodic lattice potential and the speckle potential
$U({\mathbf{r}}) = U_{L}({\mathbf{r}}) + U_{D}({\mathbf{r}})$.
The single particle Hamiltonian in units of the recoil energy
$E_{R}=\frac{\hbar^{2}k^{2}}{2m}$ is:
\begin{equation}
    {\mathcal{H}}_{1} = -\frac{\nabla^{2}}{\pi^{2}} + U({\mathbf{r}}).
\end{equation}
We constructed our potential to match the experiment as closely
as possible; see the reference \cite{demarco} for more details.

To investigate the evolution
of lattice Hamiltonian Eq.~(\ref{eq:latham2}), at every step of the imaginary time,
the basis set is orthonormalized before constructing the Hamiltonian matrix and
calculating the energies $E^{(i)}_{{\mathrm{lattice}}}$.
Then the basis set is evolved using the previous
basis set before orthonormalization; this means each basis
function is evolved independently.

\begin{figure}
\scalebox{0.3}[0.3]{\includegraphics[0,0][30cm,22cm]{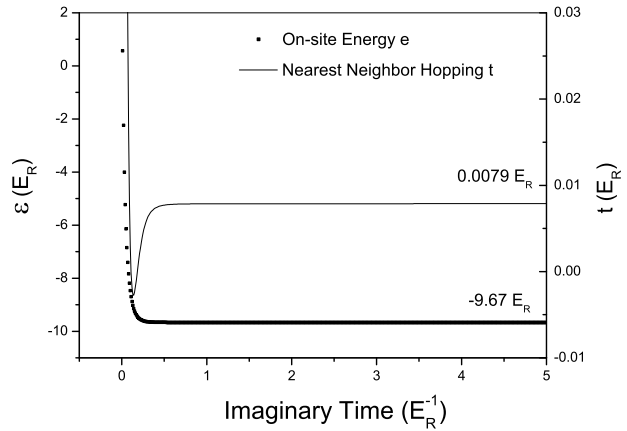}}
\caption{\label{fig:CVG1} Evolution diagram of an on-site energy(left scale) and a nearest neighbor hopping
coefficient(right scale) in a lattice for $S_{L}=14$ and $S_{D}=1$. At large imaginary time $\tau$, these two
matrix elements approach constant values.}
\end{figure}

\begin{figure}
\scalebox{0.3}[0.3]{\includegraphics[0,0][30cm,22cm]{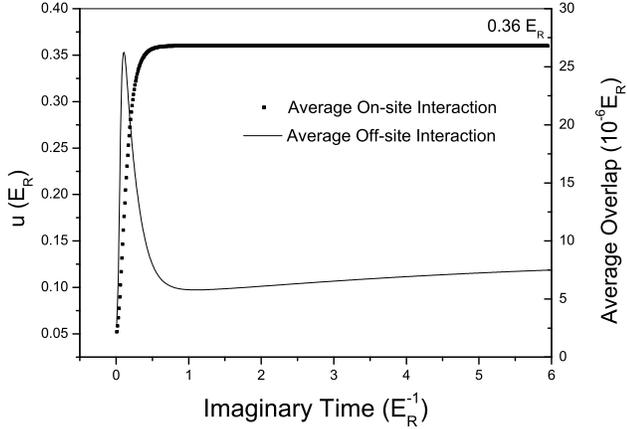}}
\caption{\label{fig:CVG2} Evolution diagram of the average on-site interaction $u$(left scale)
and the average nearest neighbor off-site interaction $\tilde{u}$(right scale) for $S_{L}=14$ and $S_{D}=1$. Note that
$\tilde{u}$ measures the localization of a pair of basis functions. The diagram shows that limiting value
of $\tilde{u}$ is $4\sim 5$ orders of magnitude smaller than that of $u$, which indicates that the basis functions
at large imaginary time are still localized.}
\end{figure}

To illustrate the convergence of the matrix elements of the lattice Hamiltonian,
the evolution diagram of an on-site energy on one particular site
and a nearest neighbor hopping coefficient on one particular bond for $S_{L}=14$ and $S_{D}=1$
are shown in Fig.~\ref{fig:CVG1}. We characterize the localization of the basis functions
by the average nearest neighbor off-site integral $\tilde{u}_{ij}$, which measures the spatial
overlap between a pair of nearest neighbor basis functions. Figs.~\ref{fig:CVG2} shows the
evolution diagrams of the average on-site interaction $u_{i}$ and the average
off-site interaction $\tilde{u}_{ij}$, which are also converging at large imaginary time. The
limiting value of the off-site interaction is $4\sim 5$ orders of magnitude smaller than that
of the on-site interaction, which means that the basis functions are still localized at large
imaginary time. Note that although the imaginary time projection operator $e^{-\tau {\mathcal{\hat{H}}}_{1}}$
spreads out the basis states, L\"{o}wdin orthogonalization operator $\hat{S}^{-1/2}$ restores
their localization form.

\begin{figure}
\scalebox{0.3}[0.3]{\includegraphics[0,0][30cm,22cm]{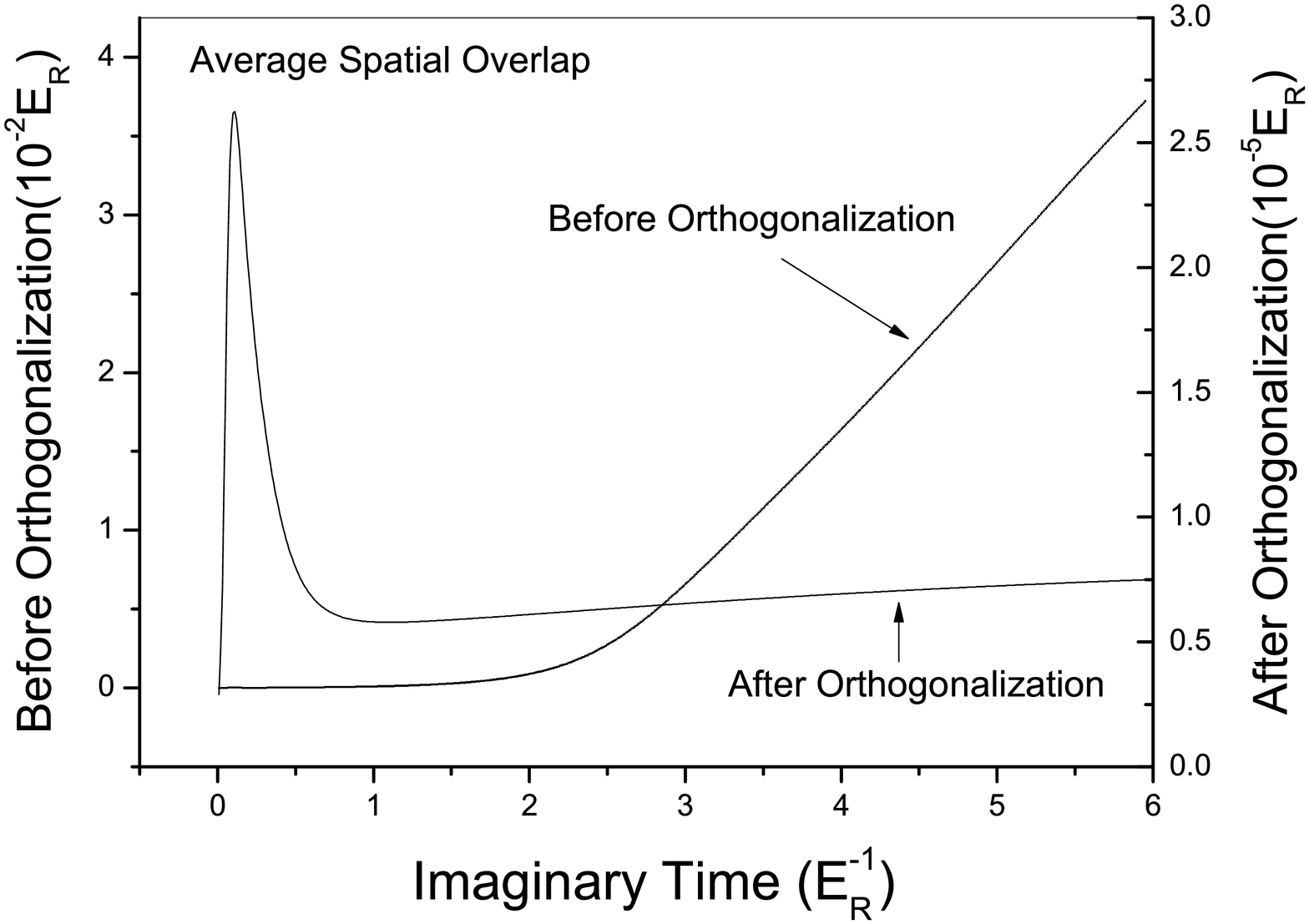}}
\caption{\label{fig:lowdin1} The effect of L\"{o}wdin
orthogonalization on the offsite integral for $S_{L}=14$ and
$S_{D}=1.0$. It can be seen from the graph that the imaginary time evolution spreads out the wave packets, but L\"{o}wdin
orthogonalization restores the localization.}
\end{figure}

To illustrate the effect of Lowdin orthogonalization on the
localization property of the basis set, the evolution diagram
for $S_{L}=14$ and $S_{D}=1$ is shown in Fig.~\ref{fig:lowdin1}
by including the off-site integral of the set before
orthogonalization. It can be seen from the graph that Lowdin
procedure helps to localize the basis functions $w(\tau)$.

\begin{figure}
\scalebox{0.27}[0.27]{\includegraphics[0,0][30cm,22cm]{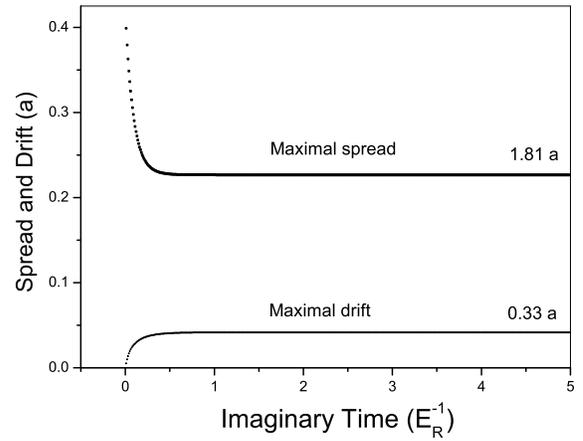}}
\caption{\label{fig:CVG3} Evolution diagram of the maximum spatial spread and drift(average deviation from the initial position)
 in units of the lattice constant for $S_{L}=14$ and $S_{D}=1$. The values that these two quantities approach at large imaginary time are small
compared to the lattice constant $a$, which means that the localization the basis functions is preserved.}
\end{figure}

The localization characterized by the spatial spread $\Omega_{w}$ and drift $D_{w}$ is shown in Fig.~\ref{fig:CVG3}.
The maximum value among all basis functions is chosen to measure the whole basis set. As shown in the graph,
the values that these two quantities asymptotes to at large time are small compared to the lattice constant, which signifies that
the basis functions are localized.

\begin{figure}
\scalebox{0.28}[0.28]{\includegraphics[0,0][30cm,22cm]{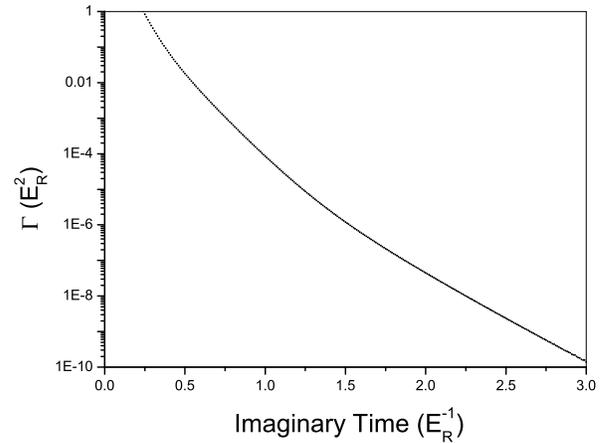}}
\caption{\label{fig:rate1} Convergence rate $\Gamma$ of eigen-energies for $S_{L}=14$ and $S_{D}=1$ shown in log-scale.}
\end{figure}

\begin{figure}
\scalebox{0.28}[0.28]{\includegraphics[0,0][28cm,20cm]{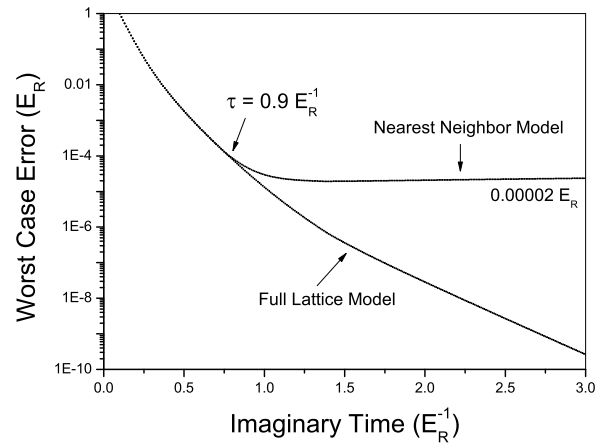}}
\caption{\label{fig:err} Imaginary time evolution of the worst case error in energy for $S_{L}=14$ and $S_{D}=1$, shown in log-scale.}
\end{figure}

The convergence rate $\Gamma$ of eigen-energies of the lattice Hamiltonian is shown in Figs.~\ref{fig:rate1}. It
can be seen from the graph that the effective lattice Hamiltonian becomes temperature-independent at low temperature. It
is also illuminating to look at the evolution diagram of the worst case error Eq.~(\ref{eq:error1}), as shown in Figs.~\ref{fig:err}
where the exact eigen-energies are estimated by
$$E^{(i)}_{\mathrm{exact}} = E^{(i)}_{\mathrm{lattice}}\left(\tau = 4E_{R}^{-1}\right).$$
We compared the worst case error in energy for the nearest
neighbor model $(\tilde{t}=0)$ versus the full lattice model. The spatial overlap between basis
functions remain negligible at the early stage so that the
nearest neighbor model has almost the same spectrum as the full
lattice model; the error in energy is reduced as imaginary time
evolves. A finite error persists in the eigen-energy of the nearest
neighbor model because the next nearest neighbor hopping terms
are neglected. Note that this error is less than $10^{-4}E_{R}$.

\begin{figure}
\scalebox{0.27}[0.27]{\includegraphics[0,0][28cm,20cm]{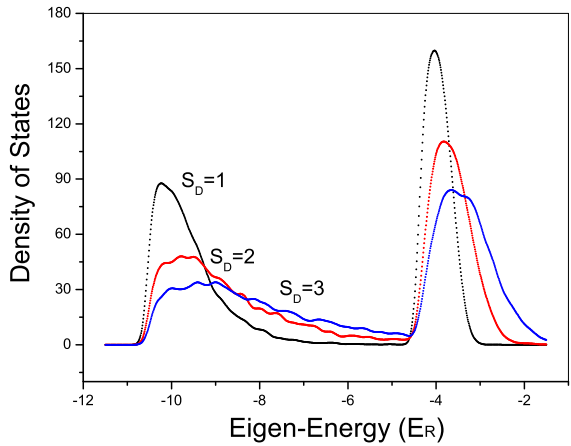}}
\caption{\label{fig:eigen1} (Color online) Density of states for a single
particle in a disordered lattice with $S_{L}=14$ and $S_{D}=1,2,3$. Energy intervals 
with low density of states still exist in the presence of disorder.}
\end{figure}

To explain how it is possible to suppress the energy of the
original localized basis set before causing significant
delocalization, it is useful to look at the single particle
density of states. In particular, we are interested in whether
the gap between bands persists in the presence of disorder.
Fig.~\ref{fig:eigen1} shows the density of states of a single
particle in the disordered lattice. 15 samples each with a
$5^{3}$ lattice for each disorder strength were calculated. It
can been seen from the plot that the lowest band is broadened
and skewed by the disorder; there remains a minimum in the
density of states between the first band and the second
band(pseudo-gap). It is the existence of such a gap that
enables us to project out the high energy components in the
initial set of trial states before delocalization sets in.

\section{Statistics of Hubbard parameters}

\begin{figure}
\scalebox{0.27}[0.27]{\includegraphics[10,0][32cm,22cm]{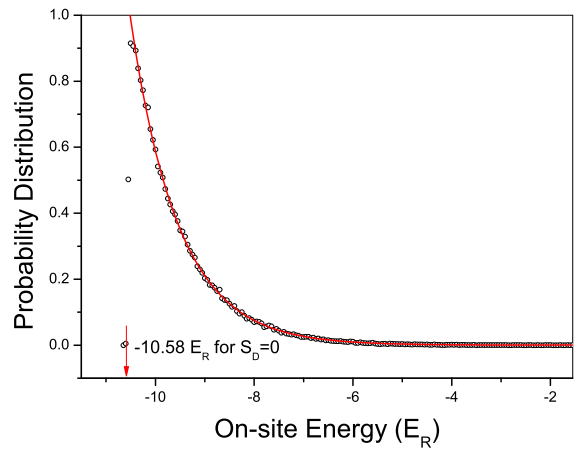}}
\caption{\label{fig:epsilon2} (Color online) Probability distribution of the
on-site energy for $S_{L}=14$ and $S_{D}=1$. The line is a fit to an exponential function.}
\end{figure}

\begin{figure}
\scalebox{0.30}[0.30]{\includegraphics[0,0][32cm,22cm]{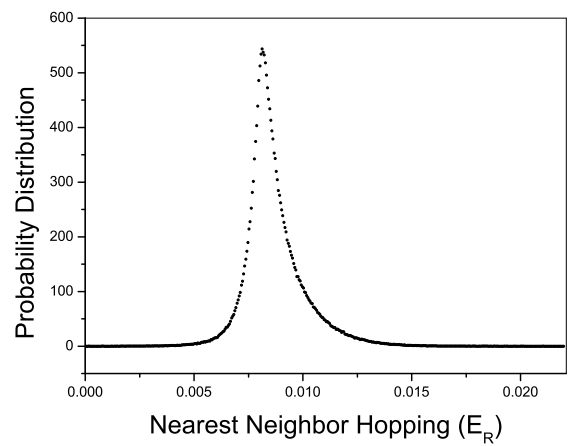}}
\caption{\label{fig:statistics1} Probability distribution of
the nearest neighbor hopping with $S_{L}=14$ and $S_{D}=1$. This is a predominantly positive asymmetric distribution.}
\end{figure}

\begin{figure}
\scalebox{0.28}[0.28]{\includegraphics[0,0][32cm,22cm]{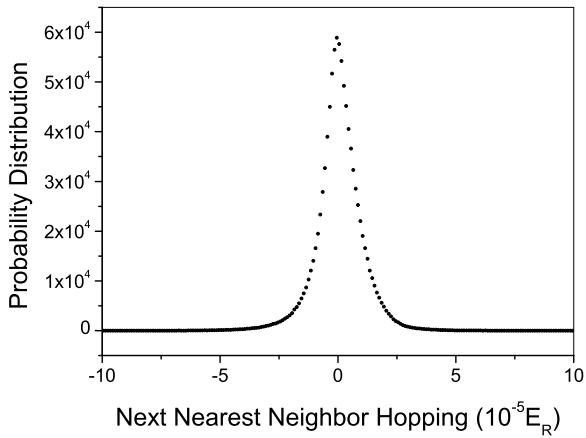}}
\caption{\label{fig:statistics2} Probability distribution of
the next nearest neighbor hopping for $S_{L}=14$ and
$S_{D}=1$. This distribution is symmetrically centered around zero.}
\end{figure}

\begin{figure}
\scalebox{0.26}[0.26]{\includegraphics[0,0][32cm,22cm]{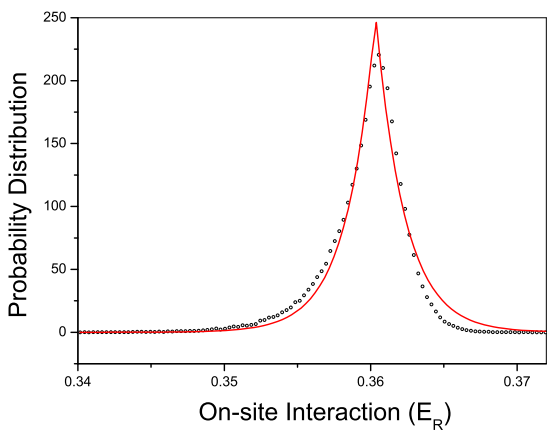}}
\caption{\label{fig:statistics3} (Color online) Probability distribution of
the on-site interaction, i.e. Hubbard U, for $S_{L}=14$ and
 $S_{D}=1$. The line is a fit to a Laplace function.}
\end{figure}

\begin{figure}
\scalebox{0.30}[0.30]{\includegraphics[0,0][32cm,22cm]{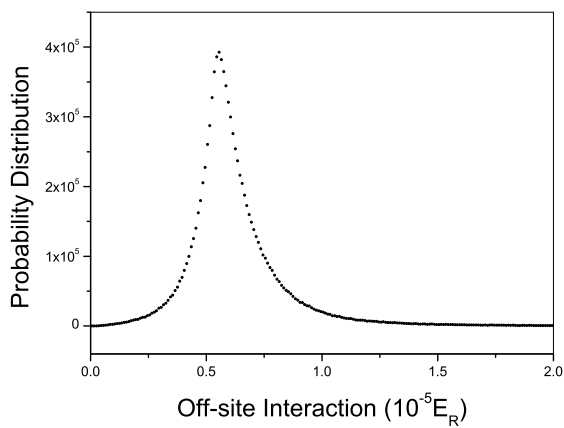}}
\caption{\label{fig:statistics4} Probability distribution of
the nearest neighbor off-site interaction for $S_{L}=14$ and
$S_{D}=1$.}
\end{figure}

We now discuss the statistical properties of the calculated
Hubbard parameters.  These are shown in
Figs.~\ref{fig:epsilon2} -~\ref{fig:statistics4} for $S_{L}=14$ and $S_{D}=1$.
About $1000$ samples of $6^{3}$ sites are used to construct the probability distributions of
Hubbard parameters.

Fig.~\ref{fig:epsilon2} shows the probability distribution of
the on-site energy $\epsilon_{i}$ for $S_{D}=1$ and
$S_{L}=14$. The distribution is skewed with a steep onset at
low energy and a tail at high energy. We fit the distribution
to an exponential decay function
\begin{equation}
    P(\varepsilon) \sim \exp\left(-\varepsilon/\Gamma\right)
\end{equation}
for $\varepsilon > -10.5E_{R}$ finding $\Gamma\approx 0.97
E_{R}$ for $S_{D}=1$ and $S_{L}=14$.  Note that the disorder
potential is always positive, so that the on site energy is
greater than its value for the periodic lattice which is $-10.58E_{R}$
for this value of $S_L$.

Hopping coefficients $t_{ij}$ characterize the mobility of the
atoms. Recall that negative values of $t$ will cause difficulty
in quantum Monte Carlo calculations. Fig.~\ref{fig:statistics1} shows the probability
distribution of nearest neighbor hopping coefficients. This
distribution is asymmetrically centered around its value $8\times 10^{-3}E_{R}$ for the
periodic potential with a width
\begin{equation}
    \frac{\delta t}{\left\langle t \right\rangle} = 0.15
\end{equation}
In $~10^{6}$ sampled bonds, only positive $t_{\langle ij\rangle}$ were found.
For $S_{L}=14$ and for $0.05\leq S_{D} \leq 1$, $\delta
t/\left\langle t \right\rangle$ ranges from $10^{-2}$ to $10^{-1}$.

Fig.~\ref{fig:statistics2} shows the probability
distribution of next-nearest neighbor hopping coefficients.
This distribution is symmetrically centered around zero with a width
$w = 1.25\times 10^{-5}E_{R}$ and
about 2 orders of magnitude smaller than nearest neighbor
hopping. Note that in the clean limit, the next nearest neighbor
hopping coefficient is exactly zero for the simple cubic lattice
by symmetry. As shown in Fig.~\ref{fig:err}, setting $\tilde{t}=0$ changes
the resulting single particle energies by a maximum of $2\times 10^{-5}E_{R}$.

Fig.~\ref{fig:statistics3} shows the probability distribution
of the Hubbard $U$, which is characterized by its narrow peak
roughly centered around the value of $u$ in the periodic limit $(0.36E_{R})$
with a $1\%$ relative width. We fit this distribution to
Laplace function
\begin{equation}
    P(u) = \frac{1}{2\Delta}\exp\left(-\frac{|u-u_{0}|}{\Delta}\right)
\end{equation}
with $u_{0}\approx 0.36E_{R}$ and $\Delta = 2\times 10^{-3}E_{R}$. For
$S_{L}=14$ and for $10^{-2}\leq S_{D} \leq 1$, $\delta
u/\left\langle u \right\rangle$ ranges from $10^{-4}$ to
$10^{-2}$. Hence one can assume that the on-site interaction is
roughly constant even in the presence of disorder.
Fig.~\ref{fig:statistics4} shows the probability distribution
of nearest neighbor overlap $u$.  We observe that the magnitude
of off-site interactions is almost 4 orders of magnitude
smaller than the on-site interaction; evidently negligible in
the many-body interacting Hamiltonian.

On-site energies are usually assumed to be almost uncorrelated
between different sites. We calculated the nearest neighbor
covariance function. For $S_{L}=14$ and $S_{D}=1$, with
$\left\langle ij\right\rangle$ nearest neighbor pairs:
\begin{equation}
\frac{\left\langle \varepsilon_{i}\varepsilon_{j} \right\rangle
- \left\langle\varepsilon_{i} \right\rangle \left\langle\varepsilon_{j} \right\rangle}{ \left\langle
\varepsilon^{2} \right\rangle - \left\langle\varepsilon
\right\rangle^{2} } \approx 0.49
\end{equation}
The $\varepsilon_{i}$'s are correlated between nearest
neighboring sites for this disordered potential.

\begin{figure}
\scalebox{0.45}[0.45]{\includegraphics[50,0][25cm,19.5cm]{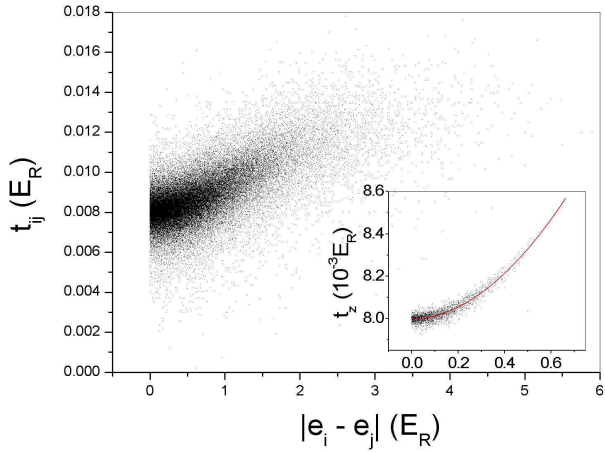}}
\caption{\label{fig:corr1} Correlation between the on-site
energy difference and hopping coefficient between nearest
neighbor sites for $S_{L}=14$ and $S_{D}=1$. The insert displays
the correlation pattern for bonds in ${\mathbf{n_{3}}}$-direction
if the longitudinal direction of the speckles is aligned with the ${\mathbf{n_{3}}}$-axis of the lattice.
The line in the insert is a fit to a power function.}
\end{figure}

Fig.~\ref{fig:corr1} shows the correlation pattern between the
on-site energy difference of nearest neighboring sites and the
hopping coefficient. Fit to this joint distribution gives
$\left\langle t_{\left\langle ij\right\rangle} \right\rangle -t_{0} \sim
\left\langle \left| \epsilon_{i}-\epsilon_{j} \right| \right\rangle^{\delta}$ with
$\delta = 1.05$. In White et. al.\cite{demarco}, the
orientation of laser speckles does not coincide with the
lattice axes; the laser speckle has a cylindrical shape whose
longitudinal direction points along
$\frac{1}{2}{\mathbf{n_{1}}}+\frac{1}{2}{\mathbf{n_{2}}}+\frac{1}{\sqrt{2}}{\mathbf{n_{3}}}$.
The insert of Fig.~\ref{fig:corr1} displays the correlation pattern for bonds in ${\mathbf{n_{3}}}$-direction
if the longitudinal direction of the speckles is aligned with the ${\mathbf{n_{3}}}$-axis of the lattice. This
illustrates the directional effect of laser speckles.

The characteristics of the speckle field is reflected in other
aspects of the parameters. According to the orientation of laser speckles
with respect to the lattice axes, we should expect that the average hopping coefficient along
${\mathbf{n_{1}}}$ and ${\mathbf{n_{2}}}$ to be equal and the
hopping along ${\mathbf{n_{3}}}$ to be different. As shown in
Table~\ref{table1},  $\langle t_{z}\rangle$ differs from those
of $\langle t_{x,y}\rangle$ because of the cylindrical symmetry
of the speckle. However, the difference is small because the
correlation length of the speckle is only
slightly larger than the lattice spacing, such that the nearest
neighbor hopping is not sensitive to the anisotropy induced by
the speckle.

\begin{center}
\begin{table}[ht]
\caption{Directional effect of speckles for $S_{L}=14$}
\begin{tabular*}{0.5\textwidth}{@{\extracolsep{\fill}} |c c c c| }

\hline

$S_{D}(E_{R})$ & $\langle t_{x}\rangle  (10^{-3}E_{R})$ & $\langle t_{y}\rangle (10^{-3}E_{R})$ & $\langle t_{z}\rangle (10^{-3}E_{R})$ \\

\hline
\hline

0.050 & $8.00\pm 2\times 10^{-4}$ & $8.00\pm 2\times 10^{-4}$ & $8.00\pm 2\times 10^{-4}$ \\

\hline

0.100 & $8.02\pm 4\times 10^{-4}$ & $8.02\pm 4\times 10^{-4}$ & $8.01\pm 3\times 10^{-4}$ \\

\hline

0.250 & $8.10\pm 1\times 10^{-3}$ & $8.10\pm 1\times 10^{-3}$ & $8.07\pm 1\times 10^{-3}$ \\

\hline

0.375 & $8.20\pm 2\times 10^{-3}$ & $8.20 \pm 2\times 10^{-3}$ & $8.16\pm 1\times 10^{-3} $ \\

\hline

0.500 & $8.32\pm 3\times 10^{-3}$ & $8.33\pm 3\times 10^{-3}$ & $8.26\pm 2\times 10^{-3}$ \\

\hline

0.750 & $8.59\pm 4\times 10^{-3}$ & $8.60\pm 4\times 10^{-3}$ & $8.48\pm 3\times 10^{-3}$ \\

\hline

1.000 & $8.72\pm 3\times 10^{-3}$ & $8.73\pm 3\times 10^{-3}$ & $8.57\pm 2\times 10^{-3}$ \\

\hline
\end{tabular*}
\label{table1}
\end{table}
\end{center}

\begin{figure}
\scalebox{0.3}[0.3]{\includegraphics[50,0][35cm,25cm]{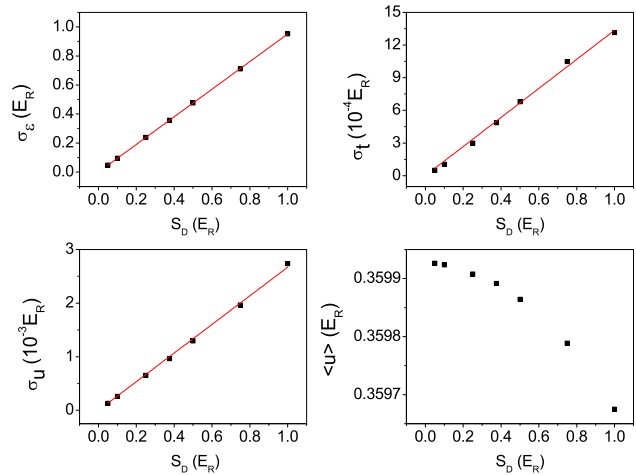}}
\caption{\label{fig:width} (Color online) The width of the probability
distribution for $\epsilon$, $t_{ij}$, $u_{i}$ and $\left\langle u \right\rangle$ for
$S_{L}=14$.}
\end{figure}

In Fig.~\ref{fig:width} the variation of the distribution
widths of all 4 Hubbard parameters versus speckle intensity is
shown for $S_{L}=14$. Fig.~\ref{fig:width}(a) shows the
dependence of the width $\sigma_{\epsilon} = \left\langle
\sqrt{\left( \epsilon_{i} - \langle\epsilon_{i} \rangle
\right)^{2}} \right\rangle$ for the onsite energy on the
disorder strength $S_{D}$ for $S_{L}=14$. It can be seen from
the graph that $\sigma$ increases linearly with the disorder
strength, approximately equal to the speckle potential shift.
Hence, the width is an appropriate measure of the disorder
strength. The linear fit of this functional dependence gives
$\sigma_{\epsilon} =  0.95\times S_{D}$. The distribution
width of nearest neighbor hopping coefficients and Hubbard $U$
are shown in Fig.~\ref{fig:width}(b) and
Fig.~\ref{fig:width}(c) respectively. In
Fig.~\ref{fig:width}(d), we show the disorder dependence of the mean value of
Hubbard $U$. It can be seen from the graph that $\left\langle u \right\rangle$ is
not sensitive to the disorder strength.

\section{Conclusion}
In this paper, we developed a method to construct low energy
basis states and applied the method to calculate the Hubbard
parameters in a disordered lattice created by an optical
speckle field.  The imaginary time projection method introduced
in this paper generates a type of Wannier-like localized basis
that satisfies several conditions imposed by a reasonable
coarse-grained, effective lattice Hamiltonian.

Detailed many-body calculations using the determined parameters
with comparison to experiments are in
progress. The method can be extended to include interactions.

\begin{acknowledgments}
We would like to thank R. M. Martin, B. L. DeMarco, F. Kruger,
M. Pasienski, B. K. Clark and Fei Lin for helpful discussions.
This work has been supported with funds from the DARPA OLE
Program.
\end{acknowledgments}

\newpage 

\end{document}